\documentstyle[multicol,aps,prl,floats]{revtex}
\begin{document}
\renewcommand{\thesection}{\arabic{section}}
\vspace{-2cm}
\draft
\title{
ANOMALOUS SCALING EXPONENTS OF A WHITE-ADVECTED PASSIVE SCALAR}
\author{M. Chertkov and G. Falkovich}
\address{
Department of Physics of Complex Systems, Weizmann Inst. of Science,
Rehovot 76100, Israel}
\maketitle
\begin{abstract}
For Kraichnan's problem of passive scalar advection by a velocity field
delta-correlated in time, the limit of large space dimensionality $d\gg1$
is considered. Scaling exponents of the scalar
field are analytically found to be 
$\zeta_{2n}=n\zeta_2-2(2-\zeta_2)n(n-1)/d$, while
those of the dissipation field are $\mu_{n}=-2(2-\zeta_2)n(n-1)/d$ 
for orders $n\ll d$.
The refined similarity hypothesis $\zeta_{2n}=n\zeta_2+\mu_{n}$ is thus 
established
by a straightforward calculation for the case considered.
\end{abstract}
\begin{multicols}{2}
It is likely that Kraichnan's model of white-advected passive scalar 
\cite{68Kra-a} will
become a paradigm in theoretical studies of
intermittency and anomalous scaling in turbulence
\cite{95KYC,95CFKL,95GK}. This is because any simultaneous
correlation function of a scalar satisfies a closed  linear differential 
equation
so that all common hypotheses about intermittency
could, in principle, be verified by direct calculation. 
In an isotropic turbulence, the $2n$-point 
correlation function depends on $n(2n-1)$ 
distances, which
makes direct solution of the respective partial differential equation 
 quite difficult in a general case. 
The fourth-order correlation function has been calculated recently
in two limiting cases: i) large space dimensionality $d\gg 1$ 
\cite{95CFKL} and ii) almost smooth scalar field $2-\zeta_2\ll1$ 
\cite{95GK}. By $\zeta_{2n}$ we designate 
the leading scaling exponent of the structure function: 
$S_{2n}(r_{12})=\langle(\theta_1-\theta_2)^{2n}\rangle
\propto r_{12}^{\zeta_{2n}}+\;$subleading terms. If $\zeta_{2n}\not=n\zeta_2$ then
the scaling is called anomalous. Anomalous scaling is the way intermittency 
manifests itself in developed turbulence: the degree of
non-Gaussianity, that may be characterized by the dimensionless ratios 
$S_{2n}/S_2^n$, depend on scale. We shall see below that the anomalous dimensions
$\Delta_{2n}=n\zeta_2-\zeta_{2n}$ are positive 
which means that the smaller the scale of fluctuations the more non-Gaussian
the statistics is.

In this Letter, we use the formalism developed in 
Ref. 3 to calculate high-order correlation functions 
assuming $1/d$ to be the smallest parameter in the problem. Following
\cite{95CFKL,95GK,95SS}, we demonstrate 
how the anomalous part of the solution appears as a zero mode with the form
independent of the pumping. Those zero modes exploit the interchange symmetry
between the distances, the number of zero modes and anomalous exponents thus
increases with the order of the correlation function.

The advection of a passive scalar
field $\theta(t,{\bf r})$ by an incompressible turbulent flow 
is governed by the equation
\begin{equation}
(\partial_t+u^{\alpha}\nabla^{\alpha}-\kappa \triangle)\theta=\phi, 
\label{a1a}
\end{equation}
$\nabla^\alpha u^\alpha=0$ being assumed.
The external velocity ${\bf u}(t,{\bf r})$ and the 
source $\phi(t,{\bf r})$ are independent random functions of $t$ and 
${\bf r}$,
both Gaussian and $\delta$-correlated in time. Their spatial characteristics
are different. The source is 
spatially correlated on a scale $L$ i.e.\ the pair 
correlation function $\langle{\phi(t_1,{\bf r}_1)\phi(t_2,{\bf r}_2)}\rangle=
\delta(t_1-t_2)\chi(r_{12})$
as a function of the argument $r_{12}\equiv|{\bf r}_1-{\bf r}_2|$
decays on the scale $L$. The value $\chi(0)=P$ 
is the production rate of $\theta^2$. The velocity field is multi-scale 
in space
with a power spectrum. The pair correlation function
$\langle  u^\alpha(t_1,{\bf r}_1) u^\beta(t_2,{\bf r}_2)\rangle=
\delta(t_1-t_2)[V_0
\delta^{\alpha\beta}-{\cal K}^{\alpha\beta}({\bf r}_{12})]$
is expressed via the so-called eddy diffusivity
\begin{eqnarray}
{\cal K}^{\alpha\beta}=\frac{D}{r^\gamma}
(r^2\delta^{\alpha\beta}-r^\alpha r^\beta)+\frac{D(d-1)}{2-\gamma}
\delta^{\alpha\beta} r^{2-\gamma}\,,\nonumber
\end{eqnarray}
where  $0<\gamma<2$ and isotropy is assumed.

Considering steady state and averaging (\ref{a1a}) 
over the statistics of ${\bf u}$ and 
$\phi$ \cite{91Kra-b,SS94}, 
one gets the closed balance equation for the simultaneous correlation
function of the scalar $F_{1\ldots 2n}=
F({\bf r}_1,\ldots,{\bf r}_{2n})\!=\!\langle\theta({\bf r}_1),
\ldots,\theta({\bf r}_{2n})\rangle$:
\begin{equation}
\!-\!\hat{\cal L}F_{1\ldots {2n}}\!\!=\!
F_{1\cdots 2n-2}\chi_{2n-1,2n}+
\!{\rm
 permutations}.
\label{eq}
\end{equation}
The operator $\hat{\cal L}\equiv \sum_{i,j}
{\cal K}^{\alpha\beta}({\bf r}_{ij})
\nabla^\alpha_{i}\nabla^\beta_{j}/2+
\kappa\sum\triangle_{i}$ describes both turbulent and molecular diffusion, 
it may be rewritten in terms of relative distances
$r_{ij}$ \cite{95CFKL}\,:
\begin{eqnarray} &&
\hat{\cal L}=\frac{D(d-1)}
{2-\gamma}\sum_{i>j} r^{1-d}_{ij}\partial_{r_{ij}} \bigl(r^{2-\gamma}_{ij}
+r_d^{2-\gamma}\bigr)r^{d-1}_{ij}\partial_{r_{ij}} 
\nonumber \\ &&
-\frac{D(d-1)}{2(2-\gamma)} \sum
(r_{in}^2-r_{ij}^2-r_{jn}^2)
\frac{r_{ij}^{1-\gamma}}{r_{jn}}
\frac{\partial^2}{\partial r_{ij}\partial r_{jn}}
\nonumber \\ &&
-\frac{D}{4}\sum
\frac{1}{r_{ij}^\gamma r_{im}r_{jn}}\biggl(
\frac{d+1-\gamma}{2-\gamma}r_{ij}^2
(r_{in}^2+r_{jm}^2-r_{ij}^2-r_{mn}^2)
\nonumber \\ &&
+\frac{1}{2}(r_{ij}^2+r_{im}^2-r_{jm}^2)
(r_{ij}^2+r_{jn}^2-r_{in}^2)\biggr)
\frac{\partial^2}{\partial r_{im}\partial r_{jn}} 
\nonumber \\ &&
+\kappa\sum \frac{r_{ij}^2+r_{im}^2-r_{mj}^2}{2r_{ij} r_{im}}
\frac{\partial^2}{\partial r_{ij}\partial r_{im}} \,. 
\label{h2} \end{eqnarray}
Here, the summation is performed over $n(2n-1)$ independent distances 
(for $d>2n-2$) with subscripts 
satisfying
the conditions $i\!\neq\! j$ and $m\neq i,j$ , $n\neq i,j$; the diffusion 
scale
$r^{2-\gamma}_d={2\kappa(2\!-\!\gamma)}/(D(d-1)$ has been introduced.
We consider the convective interval $L\gg r_{ij}\gg r_d$ where 
the operator $\hat{\cal L}$ is scale invariant.
Taking $n=1$, it is easy to find the pair correlation functions of the scalar
\cite{68Kra-a,95CFKL}\,:
\begin{equation}
F_{12}(r)=P\frac{2-\gamma}{\gamma(d-1)D}
\biggl(\frac{L^\gamma}{d-\gamma}-\frac{r^\gamma}{d}\biggr).
\label{pair}\end{equation}
We see that $\zeta_2=\gamma$.  
The scaling exponent of $\hat{\cal L}$
is $-\gamma$; the solution of (\ref{eq}) may thus be presented in the form
$F=F_{\rm forc}+{\cal Z}$, where we separated the so-called ``forced''
part of the solution (with the scaling exponent $\zeta_{2n-1}+\gamma$ prescribed
by the rhs) from the zero mode ${\cal Z}$ that may have a different scaling.  
It has been recognized
independently by the authors of \cite{95CFKL,95GK,95SS} that
they are the zero modes of the operator $\hat{\cal L}$ that are
responsible for the anomalous scaling.  We shall demonstrate below
that the factor $(L/r)^{\Delta_{2n}}$ appears in the zero mode while the factor
$(L/r)^{\Delta_{2n-2}}$ appears in the forced term,
$\Delta_{2n}$ being the anomalous dimension. The zero mode turns out to be
dominant. It has been demonstrated in
\cite{95CFKL} that $r_d$ does not appear in the leading terms of $F$
as long as at least some of the distances $r_{ij}$ are in the
convective interval. Assuming that to be the case, we omit the
diffusive parts of $\hat{\cal L}$. We shall account for them later
while considering the correlation functions of the scalar derivatives.

Let us consider now the case of large space dimensionality where the
anomalous dimensions can be calculated analytically. It is seen from
(\ref{pair}) that the level of scalar fluctuations necessary to
provide for a given flux $P$ decreases as $d$ increases.  It may be
shown that $F_{1\ldots 2n}\propto P^{n}/[d(d-1)]^{n}$. Considering
large $d$ and assuming that the flux is determined by the pumping (and
it is thus $d$-independent) we shall develop the perturbation theory
for the quantities $d^{2n}F_{1\ldots 2n}$ which have finite limits at
$d\rightarrow\infty$. We shall show below that despite the small level
of fluctuations at large $d$, the statistics of the scalar is
substantially non-Gaussian at small scales. However, since the
anomalous dimensions are small, there exists a wide interval of scales
where the correlation functions are close to their Gaussian values and
can be calculated by perturbation theory to obtain $\Delta_{2n}$ at the leading
order in $1/d$.  
As $d\rightarrow\infty$, the main part of $\hat{\cal L}$ is the operator
of the first order $\hat{\cal L}_0\!\!=\!\!(d^2{D}/(2-\gamma))
\sum\!\! r_{ij}^{1-\gamma}\partial_{r_{ij}}$.
Since $\hat{\cal L}$ is of the second order, one may wonder if the
$1/d$ perturbation theory is regular. It has been demonstrated in
\cite{95CFKL} that it is so by considering  the
perturbation theory that starts from the bare operator $\hat{\cal
L}_0'\propto\sum r^{1-d}_{ij}\partial_{r_{ij}}
r^{1+d-\gamma}_{ij}\partial_{r_{ij}}$; this alternative  theory gives the same
answer for the anomalous exponent. The difference between the
perturbation theories with $\hat{\cal L}_0$ and $\hat{\cal L}_0'$ is
in the representation for the forced term (the second approach
corresponds to some resummation). Since we are interested here in the
anomalous scaling which is given by the zero modes that are determined
by their behavior at $r=0$, there is no qualitative difference
between $\hat{\cal L}_0$ and $\hat{\cal L}_0'$ and the perturbation
theory for the anomalous exponent is regular.

The zeroth term in the perturbation series for $F_{1\ldots 2n}$ is
given by a Gaussian reducible expression. Our aim is to iterate it
once by applying the operator $\hat{\cal L}_0^{-1}(\hat{\cal
L}-\hat{\cal L}_0)$. The parameter of the expansion is $n/\gamma d$,
assumed to be small. In the first correction to be thus found, the
logarithmic terms $\ln(L/r_{ij})$ are of interest because they appear
at expanding the anomalous scaling factors $(L/r_{ij})^{\Delta_{2n}}$
over $\Delta_{2n}$ --- see \cite{95CFKL} for the details.  As we shall
see, there are different zero modes with different anomalous
dimensions $\Delta_{{2n},i}$ for any given $n$.  Of course, only the
largest $\Delta_{2n,i}$ contributes in the limit
$L/r\rightarrow\infty$. However, in the region
$\Delta_{2n,i}\ln(L/r_{ij})\ll1$, where we carry out our calculations,
all $\Delta$'s contribute logarithmic terms. We thus have a degenerate
perturbation theory (all zero modes ${\cal Z}_{2n,i}$ have the same
exponent $n\gamma$ in the zeroth approximation) and should therefore
apply the operator $\hat{\cal
L}_0^{-1}(\hat{\cal L}-\hat{\cal L}_0)$ on a vector of zero modes of
$\hat{\cal L}_0$ and  then single out the terms having logarithms.  It is
clear that logarithms may appear only multiplied by a zero mode of
$\hat{\cal L}_0$. We thus obtain the matrix of the operator $\hat{\cal
L}_0^{-1}(\hat{\cal L}-\hat{\cal L}_0)$ in the representation of
${\cal Z}_{2n,i}$. The eigenvalues of that matrix are the anomalous
exponents $\Delta_{2n,i}$ at the leading order in $1/d$.

Let us describe how the matrix is generated.
The most convenient classification of the zero modes is as follows:
${\cal Z}_{2n,i}$ is the polynomial in $x=r^\gamma$
of order $n$ which may be separated into
a symmetrical sum of polynomials, each involving distances between $i$ points.
For example, there are two zero modes for the fourth-order correlator\,:
${\cal Z}_{4,4}=\sum(x_{ij}-x_{kl})^2$ and ${\cal 
Z}_{4,3}=\sum(x_{ij}-x_{jk})^2$
\cite{95CFKL}. The first-order logarithmic correction is calculated by 
the rule
\end{multicols}
\begin{equation}
-\hat{\cal L}_0^{-1}(\hat{\cal L}-\hat{\cal L}_0) x_{ij}x_{kl}
=\frac{2-\gamma}{2d}
\ln[r]\Biggl\{
\begin{array}{c} 
(x_{ij}-x_{il})^2+\gamma (x_{ij}-x_{jl})
(x_{il}-x_{jl}),\quad i=k,\ \ j\neq l;\\
2(x_{ik}-x_{il}-x_{jk}+x_{jl})^2,
\quad  i\neq j\neq k \neq l,
\end{array}
\label{exp2}
\end{equation} 
\begin{multicols}{2}
which gives the matrix \cite{95CFKL}
\begin{equation}
\left(
\matrix{ \Delta_{4,4}& \ldots\cr 
          0 & \Delta_{4,3}\cr}\right).
\label{mat}
\end{equation}
Here, $\Delta_{4,4}=4(2-\gamma)/d$ and $\Delta_{4,3}=-(4-\gamma^2)/2d$
are the eigenvalues. Before describing the
structure of the matrix at higher $n$, let us note that the order $n$
and the number of points $i$ are not enough to specify the zero mode for
$n\geq3$ and $4\leq i\leq 2n-2$ due to the possibility of different
topological configurations (we enumerate them by $j$). 
For example, at $n=3,i=4$, there are two different zero modes: one
involves distances $r_{kl},r_{lm},r_{mn}$ and another $r_{kl},r_{ln},r_{lm}$.
The total
number of zero modes grows with $n$ faster than factorially due both to 
the growth of the number of possible functional forms and the
appearance of new topologically different configurations.  What is
important to know is that the operator acting on the zero mode with a
given $i$ produces only modes with $i'\leq i$.  The analysis of
eigenvalues is thus reduced to the consideration of the blocks with a
given $i$.  The first mode ${\cal Z}_{2n,2n}$ contains the monomial
$x_{1,2}\cdots x_{n-1,n}$ that cannot be obtained by (\ref{exp2}) from
other modes, so that the first element is $n(n-1)\Delta_{4,4}/2$ and
the remaining elements of the first column are zero. Considering ${\cal
Z}_{2n,2n-1}$, one realizes that the second diagonal element is
$(n-1)(n-2)\Delta_{4,4}/2+\Delta_{4,3}$ and all the lower elements in
the second column are zero. Then the $3\!\times\!3$ block follows
which corresponds to ${\cal Z}_{2n,2n-2,j}$:
\begin{equation}\FL\!\!\!{\Delta_{4,4}\over2}
\!\!\left(\!\matrix{{\!(n\!-\!1)(n\!-\!4})\!\!+\!4q & 0 & -2\cr
2(3-n) &{\!\!\!\!\!\!n^2\!-\!3n\!-\!1}\!-\!6q &1+4q \cr 0 & {9}+12q &
{\!\!\!\!\!\!(n\!-\!2)(n\!-\!5)}\cr}\!\right)\!\!,\label{bl3}\end{equation}
with $q=\Delta_{4,3}/\Delta_{4,4}$, all the elements below the block 
are zero.  The next block is for $i=2n-3$, it is $7\times7$ for
$n\geq6$, the sizes of the blocks grow as one approaches the center of
the matrix, then they decrease and eventually we come to the
$3\times3$ block due to ${\cal Z}_{2n,2,j}$ and a single value at the
lower right corner. By virtue of (\ref{exp2}), all entries of the
matrix at arbitrary $n$ may be expressed via those $\Delta_{4,4}$ and
$\Delta_{4,3}$ and combinatorial factors. 
Note that the classification of all eigenvalues
and eigenvectors is very important because it carries information
about the algebraic structure and underlying symmetry that governs our
set of correlation functions. We postpone the general classification
until further detailed publications.  Fortunately enough, the mode
that gives the largest anomalous dimension and the structure functions
of the dissipation field is separated so that it can be found without
finding the whole set of $\Delta_{2n,i}$.  We notice that since
$\Delta_{4,4}>
\Delta_{4,3}$ and the largest combinatorial factor in front of $\Delta_{4,4}$
appears in the first element then it is plausible to assume that
${\cal Z}_{2n,2n}$ gives the largest eigenvalue
$\Delta_{2n,2n}=n(n-1)\Delta_{4,4}/2$. One can directly check that 
$\Delta_{2n,2n-1}$ and all 
eigenvalues $\Delta_{2n,2n-2}$ of (\ref{bl3}) 
are less than $\Delta_{2n,2n}$ for any $n$.
For an arbitrary block, the validity of the assumption may be
established asymptotically for $n\gg1$ (yet, of course,
$n\ll\gamma{d}$) when all eigenvalues are $n^2\Delta_{4,4}/2+O(n)$.
For $n=2,3,4$, we found all the eigenvalues using Mathematica; the
largest is always $\Delta_{2n,2n}$.  We thus conclude that for
$\Delta_{2n,2n}\ln(L/r)\gg 1$, the main contribution to the
correlation function is given by the respective zero mode with the
scaling exponent $\zeta_{2n}=n\zeta_2-2(2-\zeta_2)n(n-1)/d$. In
particular,
\begin{equation}
\langle(\theta_1-\theta_2)^{2n}\rangle\sim r^{n\gamma}
({L}/{r})^{2n(n-1)(2-\gamma)/d}.
\label{kc}
\end{equation}
It agrees with \cite{95GK} where $\zeta_4$ has been calculated. Note that 
both (\ref{kc}) and the results of \cite{95CFKL,95GK} differ from $\zeta_{2n}$
suggested in \cite{95KYC}. In our opinion, that means that the closure implemented
in \cite{95KYC} is not exact at the limits considered ($n\ll\gamma d$ and
$2-\gamma\ll1$).

To find the correlation functions of the scalar derivatives, one 
should consider some distances $r_{ij}$ as going to zero. While
some distance passes the diffusion scale the dependence on that distance
changes. To describe that, we include the diffusion operator into
$\hat{\cal L}$. As a result, the diffusion scale
$r_d$ appears in the correlation functions. The form of $r_d$ dependence
could be readily established for an arbitrary $n,d,\gamma$ by using 
a straightforward perturbation expansion in the ratio between small and
large distances (see Sect. III of \cite{95CFKL}).
The overall scaling of $F_{1\cdots 2n}$ at all the distances 
inside the convective interval is assumed now to be known 
($\Delta_{2n}=\Delta_{2n,2n}$ for $d\gg1$)
\begin{equation}
F_{1\cdots 2n}\approx C_{2n} R^{n\gamma-\Delta_{2n}}L^{\Delta_{2n}},\quad
L\gg r_{ij}\sim R\gg r_d\ .
\label{os}
\end{equation}
Now, let us consider one distance, say $r_{12}$,
to be much smaller than the others: $\rho=r_{12}\ll r_{ij}\simeq R$.
At zero order in $\rho$,
$F_{1,1,3\cdots2n}\approx G(R)\sim R^{n\gamma-\Delta_{2n}}L^{\Delta_{2n}}$ 
\cite{95CFKL}. 
The leading isotropic correction satisfies the equation
\begin{eqnarray}&&
\hat{\cal L}_{\bbox\rho}\delta F_{2n}({\bf R},\bbox{\rho})=\hat{\cal L}_RG(R)-\Phi(R)
\label{ll}
\end{eqnarray}
 where $\Phi(R)\sim R^{(n-1)\gamma-\Delta_{2n-2}}L^{\Delta_{2n-2}}$ 
is the major ($R$-dependent) 
term of the rhs of (\ref{eq}),
$\hat{\cal L}_R$ is the major term of the operator $\hat{\cal L}$,
and $\hat{\cal L}_{\bbox\rho}$ is the perturbation operator 
$\hat{\cal L}_{\bbox\rho}\equiv
{\cal K}^{\alpha\beta}({\bbox\rho})
\nabla_{\rho}^\alpha\nabla_{\rho}^\beta
+2\kappa\triangle_{\rho},\label{Lrho}$. The solution has the form
\begin{equation}
\delta F_{2n}({\bf R},\bbox{\rho})\sim R^{(n-1)\gamma-\Delta_{2n}}L^{\Delta_{2n}} 
\int^{\rho}_0
\frac{rdr}{2r_d^{2-\gamma}+r^{2-\gamma}},
\label{cor}\end{equation}
In deriving (\ref{cor}), it has been implied that $\Delta_{2n}>\Delta_{2n-2}$.
At $\rho\ll r_d$, the isotropic correction is analytic in $\rho$:
$\delta F_{2n}({\bf R},\bbox{\rho})\sim R^{(n-1)\gamma-\Delta_{2n}}L^{\Delta_{2n}}
r_d^{\gamma-2}\rho^2$. Now we are ready to differentiate it with respect to
$\rho$, in particular, to calculate the correlation functions that involve
the dissipation field $\epsilon=\kappa[\nabla\theta]^2$:
\begin{equation}
\langle\langle\epsilon_1\theta_3\cdots\theta_{2n}\rangle\rangle
\sim R^{(n-1)\gamma}\bigl({L}/{R}\bigr)^{\Delta_{2n}}\ .
\label{ethpa}\end{equation}
Since $\epsilon\propto \kappa\propto r_d^{2-\gamma}$ then $r_d$
disappears.
The diffusion scale appears only at the anisotropic
terms proportional to angular harmonics with respect to the angle 
between ${\bf R}$ and 
$\bbox{\rho}$. This is because of the zero modes of the operator 
$\hat{\cal L}_{\bbox\rho}$ associated with
the angular harmonics that are expressed via the Jacobi polynomials 
$P_{2k}^{(\nu,\nu)}$ \cite{95CFKL,95proc}:
\begin{eqnarray}&&
{\cal Z}_{2k}(\bbox{\rho},{\bf R})=
P_{2k}^{(\nu,\nu)}\biggl[\frac{{\bf n}\bbox{\rho}}{\rho}\biggr]\times
\biggl\{\begin{array}{c}\rho^{\delta_k},\quad \rho\gg r_d,\\
\rho^{2k}r_d^{\delta_k-2k},\quad \rho\ll r_d;\end{array},
\nonumber
\\&&
\delta_k\!=\! \frac{1}{2}\!
\left(\gamma\!-\!d\!+\!\sqrt{
(d-\gamma)^2\!+\!\frac{8k(d+1-\gamma)(2k+d-2)}{d-1}}\right),
\nonumber
\end{eqnarray} 
where $\nu=(d-3)/2$, ${\bf n}={\bf R}/R$. 
The $2k$-th angular correction 
to $F_{1,1,3,\cdots,2n}$ is thus
\begin{equation}
\delta^{(k)} F_{2n}(\bbox{\rho},{\bf R})\sim {\cal Z}_{2k}(\bbox{\rho},{\bf R})
\bigl({L}/{R}\bigr)^{\Delta_{2n}} R^{n\gamma-\delta_k},
\label{nharm}
\end{equation}
where the leading scaling behavior of the $\rho$-independent prefactor
is restored to give a correct dimensionality. 
The overall ($L$-dependent) scaling is assumed to be known.
If $n\gamma>\delta_k$ and $\rho$ is in the diffusive interval, then
the estimate (\ref{nharm}) 
is true only at $R$ being small enough: 
$(R/L)^{\Delta_{2n}}(r_d/R)^{n\gamma-\delta_k}\ll 1$ --- see
\cite{95CFKL} for the details. Differentiating $\delta^{(2k)}F$
with respect to $\bbox{\rho}$ one can establish the scaling of the correlation
functions that involve high-order traceless tensors of derivatives
$\xi^{(k)}=\kappa^{(2k-\gamma)/(2-\gamma)}\sum_{l=0}^k a_l
\bigl[({\bf n}\nabla)^{k-l}\nabla^{\alpha_1}\cdots\nabla^{\alpha_l}\theta\bigr]^2$,
$a_l$ is a respective coefficient of $P_{2k}^{(\nu,\nu)}(x)$ expansion
in the series in $x$.
One can also generalize (\ref{ethpa}) in another way, considering extra fusion of
another one or more pairs of points.
The results show that the fusion procedures for an arbitrary
number of different pairs of points commute with each other.
In other words, using the language of the so-called
operator algebra \cite{69Pol,69Kad,69Wil} 
the validity of which for turbulence was argued in \cite{93Eyi,94LL},
we can introduce $r_d$-related dimensionality of the fields:
the scalar field $\theta$ and dissipation field $\epsilon$ have dimensionality $0$, 
while $\xi^{(k)}$
has dimensionality $\delta_k-\gamma$.
Thus, to find the ultraviolet ($r_d$-related) dimensionality
of a composite multiplicative field one should sum the dimensionalities
of the multipliers:
$$\Bigl\langle\theta_1\cdots\theta_l\epsilon_{l+1}\cdots\epsilon_m
\prod_k[\xi^{(k)}]^{c_k}\Bigr\rangle
\propto {r_d}^{\sum c_k(\gamma-\delta_k)}\ .$$

The above fusion rules describe the scaling of the fluctuations
of the enumerated primary fields only. 
Particularly, the rules give the scaling of 
correlation functions with $\epsilon$ at different points 
but not with the higher powers of the dissipation field,
$\epsilon^n$. The difference stems from the fact that to calculate
the correlation functions containing $\epsilon^n$ we should fuse a group
of $2n$ (rather than two)  points in an initial $\theta$ correlator.
To calculate a
correction that produces a nonzero contribution into
the desirable correlator of $\epsilon^n$, one should compare
the leading forced correction with a zero mode of the $2n$-point
operator, but not with that of two-point one $\hat{\cal L}_{\bbox\rho}$.
Such zero modes do have anomalous scaling, as we have learned above,
they produce the major contribution to correlation functions
of $\epsilon^n$. The ultraviolet anomalous scaling of such objects 
is thus related to the infrared anomalous scaling of the passive scalar itself.
For example, by a direct application of the above procedure (\ref{ll}-\ref{ethpa}),
we get
\begin{equation}
\langle\epsilon_1^n\epsilon_2^m\rangle\!\sim\!
\left({L}/{r_{12}}\right)^{\Delta_{2n+2m}-\Delta_{2n}-\Delta_{2m}}\!
\left({L}/{r_d}\right)^{\Delta_{2n}+\Delta_{2m}}.\label{uu}\end{equation}
The dissipation field is thus highly intermittent, the single-point means
$\langle\epsilon^n\rangle\sim \langle\epsilon\rangle^n(L/r_d)^{\Delta_{2n}}$
grow unlimited when diffusivity decreases. Statistics at the convective interval
is better related to local average $\epsilon_r$ over the
ball with the radius $r$. Since spatial integration and time average commute
then our knowledge of 
$\langle\epsilon_1\cdots\epsilon_n\rangle$ with all distances in the convective
interval allows one to obtain by spatial integration (which converges if 
$\Delta_4<d$):
\begin{equation}\FL
\!\!\langle(\epsilon_r)^n\rangle\sim \langle\epsilon\rangle^{n}
(r/{L})^{\mu_n},\  \mu_n=\zeta_{2n}-n\zeta_2=-\Delta_{2n}\ .
\label{ka}
\end{equation}
That relation presents a version of the refined similarity hypothesis 
\cite{62Kol,62Oby,95SKS} valid in our case. Indeed,
scalar field and dissipation field are related: 
$\epsilon_r\!\simeq\!(\delta\theta_r)^2/\tau_r$ where
$\tau_r\!\simeq\! Dr^{\gamma}$ is a transfer time. For $n=2$, (\ref{ka}) 
has been established in \cite{95KYC,95CFKL}.

The first ($n\!\ll\! \gamma{d}$) moments of the locally averaged dissipation
have $\mu_n=-n(n-1)\Delta_{4,4}/2$
so they are described by $log$-normal 
statistics \cite{41Kold}\,:
$${\cal P}(\epsilon_r)\sim
\exp\biggl(-\frac{(\ln[\epsilon_r/\langle\epsilon\rangle]-
\Delta_{4,4}\ln[L/r]/2)^2}
{2\Delta_{4,4} \ln[L/r]}\biggr)\ .$$

To conclude, the law (\ref{kc}) qualitatively corresponds to the observed
behavior  of $\zeta_n$ \cite{84AHGA,90MS,91Sre}. Quantitatively, we cannot
use (\ref{kc}) for $d=3,\gamma=2/3$ (which would give an 
overestimation of the anomalous exponents)  because the validity condition
of our theory $n\ll\gamma d$ will be violated already at $n=2$.
Note that quadratic dependence of $\zeta_n$ and $\mu_n$ on $n$
(and lognormality) is violated
when  $n\simeq\gamma{d}$, while the similarity relation (\ref{ka}) 
is true for any $n,\gamma,d$.

This work is a part of the extensive program on studying anomalous
scaling in turbulence undertaken together with E. Balkovsky, I. Kolokolov
and V. Lebedev. We are grateful to them for numerous discussions.
We are indebted to G. Eyink, U. Frisch, R. Kraichnan and the 
anonymous referee for useful remarks. 
The work was supported by the Clore Foundation (M.C.) and by the Rashi Foundation 
(G.F.). 

\end{multicols}
\end{document}